\documentclass[prb,8pt,twocolumn,superscriptaddress]{revtex4-1}
\usepackage{amssymb,graphicx,color,amsmath,xfrac, oldstyle,bm,ifthen,lineno}
\usepackage[all]{xy}

\usepackage[mathletters]{ucs}
\usepackage[utf8x]{inputenc}

\definecolor{lightblue}{rgb}{0.17,0.39,1}
\definecolor{lightgreen}{rgb}{0.67,0.81,0.08}
\definecolor{lightred}{rgb}{1,0.05,0.52}
\usepackage[bookmarks, colorlinks=true, breaklinks, pdftitle={On the amplitude of quantum oscillations in correlated metals}, pdfauthor={Arkady Shekhter}]{hyperref}
\hypersetup{linkcolor=black,citecolor=black,filecolor=black,urlcolor=black}

\newcommand{\YBCO}[1]{$\text{YBa}_2\text{Cu}_3\text{O}_{#1}$}
\newcommand{\BaAsPO}{$\text{BaFe}_2(\text{As}_{1-x}\text{P}_x)_2$}

\newcommand{\matzln}{{\overline{\text{ln} }}}

\newcommand{\HT}[1]{{\mathrm{H.T.}\left\{#1\right\} }}

\newcommand{\Q}{{\mathcal{Q}}}

\newcommand{\SE}{{\hat{Σ}}}
\newcommand{\sign}{{\mathrm{sign}}}

\renewcommand{\Re}{{\mathrm{Re}}}
\renewcommand{\Im}{{\mathrm{Im}}}
\renewcommand{\tanh}{{\mathrm{tanh}}}

\renewcommand{\t}[1]{\ensuremath{\text{#1}}}

\newcommand{\f}[2]{\frac{#1}{#2}}

\newcommand{\para}[1]{\left(#1\right)}
\newcommand{\brac}[1]{\left[#1\right]}

\newcommand{\bg}{\t{bg}}

\newcommand{\osc}{\t{osc}}

\newcommand{\stat}{\t{stat}}
\newcommand{\dyn}{\t{dyn}}
\newcommand{\mfrac}[2]{\!\para{#1/#2}}
\newcommand{\liminff}{\limits_{-∞}^{∞}\!\!\!}

\SelectTips{eu}{12}

\newcommand{\fancy}[1]{{#1}}
\newcommand{\x}{\object@{>}}
\newcommand{\y}{\object@{<}}
\newcommand{\s}[1]{{\scriptscriptstyle{#1}}}
\newcommand{\tadpole}[4]{{{ \fancy{\xymatrix @=0.7pc @M=0pt @C=1.2pc{
 \ar@{-}[r]|{\x}_{\s{#1}}  & \ar@/^1.1pc/@{~}[rr]^{\s{#2}} \ar@{-}[rr]|{\x}_{\s{#3}} &&  \ar@{-}[r]|{\x}_{\s{#4}}&}}}}}
\begin{document}
\title{Thermodynamic constraints on the amplitude of quantum oscillations.}

\author{Arkady~Shekhter}
\affiliation{National High Magnetic Field Laboratory, Florida State University, Tallahassee, Florida 32310, USA}

\author{K.~A. Modic}
\affiliation{Max-Planck-Institute for Chemical Physics of Solids, Noethnitzer Strasse 40, D-01187, Dresden, Germany}

\author{R.~D. McDonald}
\affiliation{Los Alamos National Laboratory, Los Alamos, New Mexico 87545, USA}

\author{B.~J. Ramshaw}
\affiliation{Los Alamos National Laboratory, Los Alamos, New Mexico 87545, USA}
\affiliation{Laboratory for Atomic and Solid State Physics, Cornell University, Ithaca, New York 14853, USA}

\begin{abstract}
Magneto-quantum oscillation experiments in high temperature superconductors show a strong thermally-induced suppression of the oscillation amplitude approaching critical dopings\cite{Ramshaw-Science2015, Shishido-PRL2010,Walmsley2014}---in support of a quantum critical origin of their phase diagrams. We suggest that, in addition to a thermodynamic mass enhancement, these experiments may directly indicate the increasing role of quantum fluctuations that suppress the oscillation amplitude through inelastic scattering. We show that the traditional theoretical approaches beyond Lifshitz-Kosevich\cite{Lifshitz-Kosevich} to calculate the oscillation amplitude in correlated metals result in a contradiction with the third law of thermodynamics and suggest a way to rectify this problem.
\end{abstract}
\date{\today}\maketitle
\setlength{\parindent}{0.5cm}
\setlength{\parskip}{0.5cm}

Recent advances in high-magnetic field measurements have amassed a body of information about metallic quantum criticality in high-temperature superconductors.\cite{Mackenzie2013, Fisher2014, Hussey2009, Zaanen-review-2015} In particular, quantum oscillation measurements in cuprate, \YBCO{6+x}\cite{Ramshaw-Science2015}, and pnictide, \BaAsPO\cite{Shishido-PRL2010,Walmsley2014}, systems suggest a strong enhancement of the quasiparticle mass approaching a critical doping---a locus for thermodynamic anomalies in other measurements as well \cite{Teillefer,Loram,Walmsley2014}.

The quasiparticle mass, $m^{⋆}$, is inferred in these measurements from analysis of the temperature dependence of the quantum oscillation amplitude $A(T)$.\cite{Shoenberg} In conventional metals, $A(T)$ decays over a temperature range that is inversely proportional to the quasiparticle mass with no parameters other than the magnetic field and temperature entering the functional form, $A_0(T)=X/\sinh{X}$, where $X=2π^2(k_BT)/(ℏω_c)$ and $ω_c=eB/m^{⋆}$ is the cyclotron frequency.\cite{Lifshitz-Kosevich} This form of the temperature dependence of $A(T)$ originates in the temperature smearing (over an energy interval $k_BT$) of the occupation number of Landau levels (spaced at $ℏω_c$) near the Fermi surface. Importantly, it relies on the presence of well-defined quasiparticles near the Fermi surface, justified by the Fermi liquid theory of conventional metals.\cite{Luttinger-QO,Luttinger}

Unlike the renormalizations of $m^{⋆}$, which describe changes in electron velocity without changes in lifetime, electron interactions in correlated metals lead to anomalous quasiparticle relaxation dynamics, observed via the temperature and energy dependence of the quasiparticle relaxation rate $1/τ(T,ϵ)$.\cite{Mackenzie2013, Fisher2014, Hussey2009, Zaanen-review-2015} Such a departure from Fermi liquid behavior must change the character of quantum oscillations, or at the very least add to the temperature dependence of $A(T)$ and change its interpretation in terms of the quasiparticle mass.

\begin{figure}[ttt!!!!]
\centerline{\includegraphics[width=\columnwidth]{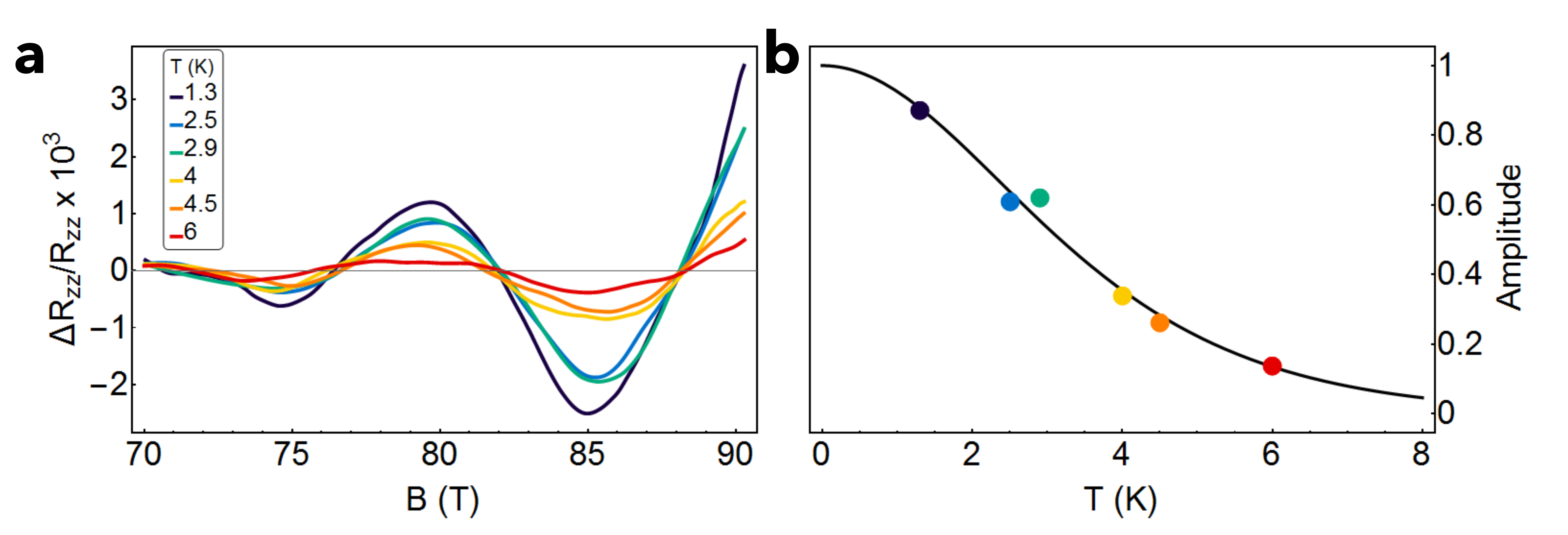}}
\caption{The temperature dependence of quantum oscillations in \YBCO{6+x} near optimal doping. \textbf{a} Quantum oscillations in the c-axis resistivity, $ρ_{zz}$, up to 90T for \YBCO{6.86} ($p=0.152$, $T_c=92K$). \cite{Ramshaw-Science2015} \textbf{b} The temperature-dependent amplitude of the oscillations in \textbf{a} closely follows the standard Lifshitz-Kosevich behavior $A(T)$ (black line). }
\label{fig:data}
\end{figure}

It is therefore puzzling that the observed temperature dependence of $A(T)$ in high-temperature superconductors (Figure~\ref{fig:data}) appears, within experimental resolution, to be identical to its Lifshitz-Kosevich form $A_0(T)$, even for chemical compositions near the critical doping\cite{Ramshaw-Science2015, Shishido-PRL2010,Walmsley2014} where strong correlation effects are well established.\cite{Mackenzie2013, Fisher2014, Hussey2009, Zaanen-review-2015} Herein we discuss the thermodynamic constraints on the form of $A(T)$ that derive directly from the third law of thermodynamics (Nernst's theorem) and are independent of the nature of the metallic state. Importantly, the form of $A(T )$ required by these constraints suggests that the observed $A(T)$ is universal in its broad features that are common to Lifshitz-Kosevich form, namely, vanishing slope at zero temperatures and monotonic decay as temperature is increased.

Whether the doping evolution of $A(T)$ observed in Refs.~\onlinecite{Ramshaw-Science2015, Shishido-PRL2010,Walmsley2014} approaching critical doping is a result of quasiparticle mass evolution alone is now an open question. More insight into the dynamic and static effects near a quantum critical point can be gained from comparison between masses obtained in quantum oscillation measurements with those from other experimental probes, such as heat capacity, cyclotron resonance, and Landau level spectroscopy.

At low temperatures, metals in a magnetic field exhibit quantum oscillations---an oscillatory variation of magnetization, resistivity, and other properties with field intensity.\cite{Shoenberg} The frequency, $F$, of the oscillations, which are periodic in inverse magnetic field, $1/B$, has a direct geometric interpretation as the extremal area (perpendicular to the field) of the Fermi surface in momentum space.\cite{Onsager1952} Quasiparticle properties near the Fermi surface, such as the effective mass $m^{⋆}$ and the relaxation time $τ$, can be obtained from the analysis of the field and temperature dependences of the oscillation amplitude $A(T,B)$. The amplitude $A(T,B)$ is defined via the oscillatory part of the thermodynamic potential, $Ω_\osc(T,B)∝A(T,B)\cos(2πF/B+γ)$. Note that $A(T,B)$ contains all of the temperature dependence in this expression, therefore, it determines the oscillatory part of entropy, $S_\osc(T,B)=-∂Ω_\osc(T,B)/∂{T}$.

Nernst's theorem requires that the {total} entropy vanishes in the zero-temperature limit, $[S_\osc(T,B)+S_\bg(T,B)]_{T\!\!→0}=0$, where $S_\bg(T,B)$ is the non-oscillatory part.\cite{Callen1960,Finkelstein-heat} This requirement can only be satisfied at all fields if $S_\osc(T,B)$ itself vanishes in the zero temperature limit because of its distinct magnetic field dependence, $S_\osc(T→0,B)=0$. The temperature derivative of the amplitude of quantum oscillations must therefore vanish in the zero-temperature limit,
\begin{align}\label{eq:slope}
\para{\frac{∂A(T,B)}{∂T}}_{T→0}=0\,.
\end{align}
Similarly, the oscillating part of the heat capacity, $c_\osc(T,B)=T(∂S_\osc(T,B)/∂T)$, must vanish at zero temperature. This requires that the curvature of $A(T,B)$ is less singular than $1/T$ in the limit of zero temperature,
\begin{align}\label{eq:curvature}
\para{T\frac{∂^2A(T,B)}{∂T^2}}_{T→0}=0\,.
\end{align}
We emphasize that this line of reasoning does not provide justification for the presence of quantum oscillations in any metal, but rather it sets tight bounds on the behavior of $A(T,B)$ if the metal does exhibit quantum oscillations in the zero-temperature limit, i.e., if $A(T,B)$ can be defined at all. It is in this sense that these thermodynamic identities are independent of the character of the metallic state.

Being a result of a microscopic calculation, the Lifshitz-Kosevich $A_0(T)$ satisfies identically both Eqs.~(\ref{eq:slope})~and~(\ref{eq:curvature}): $A_0(T)$ approaches zero temperature with zero slope and finite (non-singular) curvature. The temperature dependence of $A_0(T)$ in a broad temperature range is tightly constrained by requirements of vanishing entropy at zero temperature, $dA_0(T)/dT|_{T→0}=0$, and fast decay at high temperatures. Furthermore, its temperature dependence is set by a single energy scale, $ℏω_c$, which requires $A_0(T) = f_0(X)$, where the function $f_0(X)$ approaches zero with zero slope and decays for $X≳1$.

The form of $A(T)$ in the quantum-critical regime need not be identical to that of LK because the underlying metallic state is not the same as the one assumed in calculation of the Lifshitz-Kosevich form of temperature dependence. However, for the scale invariant dynamics\cite{Zaanen-review-2015}  in the vicinity of the quantum critical point, the temperature dependence of $A(T)$ can only be set by an energy scale associated with an external magnetic field, $α×ℏω_c$. Thus, $A(T) = f_q(X/α)$ where $α$ is a numeric factor. The function $f_q(X)$ is similar in its form to $f_0(X)$: It must approach zero temperature with zero slope (Nernst's theorem), it must decay for $X≳1$ and it has no anomalies other then smooth crossover around $X\approx1$. Further refinement in microscopic modeling and experimental sensitivity may point to more subtle differences in the functional form of the two temperature dependences. Unlike the electron renormalization in $m^{⋆}$ which describes changes in electron velocity without changes in its lifetime, $α$ originates in quasiparticle relaxation dynamics, via the temperature and energy dependence of the quasiparticle relaxation rate $1/τ(T,ϵ)$. We note that fitting $A(T) =f_q(X/α)$ with $A_0(T) =f_0(X)$ yields $m^{⋆}/α$ rather then the quasiparticle mass $m^{⋆}$.\cite{doping}

Thus, the observed Lifshitz-Kosevich functional form of the temperature dependence of the amplitude of quantum oscillations in high-temperature superconductors\cite{Ramshaw-Science2015, Shishido-PRL2010,Walmsley2014} should not be taken as conclusive evidence of Fermi liquid behavior, where all electron scattering is elastic and interaction effects are captured by mass renormalization.

The existing theoretical discussions of the amplitude of quantum oscillations in correlated metals all lead to finite entropy at zero temperature, $dA(T)/dT|_{T→0}≠0$ \cite{Wasserman-Rep, SteveJulian, Hartnoll, Pelzer1991, MyakeVarma, WassermanKarniewicz, WassermanBallesil, Riseborough,one}, violating Nernst's theorem. In the remainder of this note we attempt to identify the source of the problem and its resolution. In particular, we suggest that rather than signaling the inadequacy of their common starting point (Luttinger's functional representation of the Free energy \cite{Luttinger}), the violation of Nernst's theorem in these discussions simply indicates that the approximation scheme chosen is inconsistent, following too closely the approximation scheme used for the Fermi liquid metal.

The standard starting point for the existing theoretical analysis of quantum oscillations in a strongly interacting electron liquid\cite{Wasserman-Rep, SteveJulian, Hartnoll, Pelzer1991, MyakeVarma, WassermanKarniewicz, WassermanBallesil, Riseborough} is the Luttinger's functional representation for the thermodynamic potential \cite{Luttinger,LeeYang},
\begin{align}
Ω = T∑_{iε_n,p}\ln\!\para{G} +Y\{G\} -T∑_{iε_n,p} G(G_0^{-1}-G^{-1})\,,
\label{eq:LF}
\end{align}
where $G(iε_n,p)=1/[iε_n-ϵ(p)+μ-Σ(iε_n,p)]$ is the exact Green's function for interacting electrons, and $G_0(iε_n,p)=1/[iε_n-ϵ(p)+μ]$ is the Green's function for free electron propagation.\cite{Negele-Orland} $Y\{G\}$ stands for an infinite set of diagrams in which electron quasiparticle propagation is represented by $G(iε_n,p)$\cite{Luttinger}.

The first term in Eq.~(\ref{eq:LF}) evaluates to
\begin{align}\label{eq:omega-one}
Ω_1\!=\!-\!∫\f{dε}{2π}\tanh\f{ε}{2T}∑_p\Im\,\matzln\brac{ε\!-\!ϵ_p\!+\!μ\!-\!Σ^R(ε,T)}\!,
\end{align}
where $ϵ_p$ are the energies of quasiparticle states. This term is similar in structure to Green's function representation of the thermodynamic potential for the non-interacting Fermi gas, $Ω_0= T∑_{iε_n,p}\ln\!\para{G_0}=-T∑_p\ln\!\!\para{1+e^{\f{μ-E_p}{T}}}$ (see Appendix for details). In Fermi liquid metals, the oscillatory part of the thermodynamic potential is obtained by analysis of this term, $Ω_1$, alone.\cite{Luttinger-QO}

The single-particle energy levels $ϵ_p$ in a magnetic field consist of a set of quantized Landau levels $λ_p=ω_c(p+1/2)$ with degeneracy $BA/Φ_0$, where $A$ is the area of the crystal, $Φ_0$ is the flux quantum, $ω_c=eB/m^{⋆}$ is the cyclotron frequency, and $m^{⋆} = \para{\sfrac{ℏ^2}{2 π} }\para{\sfrac{∂ A_p}{∂ϵ}}$. $A_p$ is the Fermi surface area perpendicular to the magnetic field.\cite{Lifshitz-Kosevich,Luttinger-QO} Performing the sum over the Landau level index $p$ in Eq.~(\ref{eq:omega-one})  (see Appendix for details) we obtain for the oscillatory (periodic in $2π/ω_c$) part of $Ω_1$ per unit area,
\begin{align} \label{eq:omega-one-after}
Ω_{1,\osc} =& - 2mϖ_c ∫\liminff\f{dε}{2π}\tanh\para{\f{ε}{2T}}\notag\\
&\qquad×\Im\,\matzln\para{1+ e^{-i\f{ε-Σ^R(ε,T)+μ}{ϖ_c}}}\,,
\end{align}
where $ϖ_c=ω_c/2π$. Quasiparticle lifetime effects are introduced in Eq.~(\ref{eq:omega-one-after}) via the temperature and energy dependence of the imaginary part of the self-energy, $1/τ=-2\ImΣ^R(ε,T)$. The real part of the self-energy, $\ReΣ^R(ε,T)$, is responsible for quasiparticle mass renormalization\cite{Negele-Orland}. We distinguish the ``static" and ``dynamic" parts of the quasiparticle relaxation rate, $\ImΣ^R(ε,T)=\ImΣ^R_\stat + \ImΣ^R_\dyn(ε,T)$. The static part, $\sfrac1{τ_0}=-2\ImΣ^R_\stat$, is independent of temperature and energy, equal to $\ImΣ^R(ε→0,T→0)$. This term typically describes the effects of elastic disorder, and introduces a temperature-independent exponential envelope in field to the oscillation amplitude---the so-called ``Dingle'' factor.\cite{Shoenberg} The dynamic part, $\sfrac1{τ_\dyn} = -2\ImΣ^R_\dyn$, left out of calculations of $A_0(T)$ in the Fermi liquid,\cite{Shoenberg} contains all the temperature and energy dependence of $\ImΣ^R(ε,T)$ and is constrained by $\ImΣ^R_\dyn(ε→0,T→0)=0$. For the remainder of this discussion we will explicitly focus on the dynamic effects introduced through $\ImΣ^R_\dyn$, and all mass-renormalization effects that enter through $\ReΣ^R(ε,T)$ are contained in the renormalized $ϖ_c$.

The principal harmonic of $Ω_{1,\osc}$, Eq.~(\ref{eq:omega-one-after}), defines the temperature dependent amplitude $A_1(T)$
\begin{align}\label{eq:AKcorr}
Ω_{1,\osc} ∝& R_DA_1(T)\;\cos\para{\sfrac{μ}{ϖ_c}},  \notag\\
A_1(T)&=\frac{1}{ϖ_c}∫\liminff\frac{dε}{2}\sin\!\para{\frac{ε}{ϖ_c}}\tanh\!\para{\frac{ε}{2T}}\;e^{-\frac{\ImΣ^R_\dyn(ε,T)}{ϖ_c}}\,,
\end{align}
where the temperature-independent Dingle factor $R_D=e^{-\sfrac1{2ϖ_cτ_0}}$ accounts for the effects of elastic scattering. In the limit where the dynamic part of the self-energy vanishes, $\ImΣ^R_\dyn(ε,T)=0$, $A_1(T)$ reduces to its Fermi liquid form, $A_0(T)=-\para{\sfrac{1}{ϖ_c}}∫dε\sin\para{\sfrac{ε}{ϖ_c}}n_F(ε)=X/\sinh{X}$, where $X=πT/ϖ_c$ and $n_F(ϵ)=\brac{1-\tanh\sfrac{ε}{2T}}/2$   is the Fermi-Dirac distribution function. This limit is used to set the normalization factor for $A_1(T)$ in Eq.~(\ref{eq:AKcorr}). Unless $A_1(T)$ satisfies the thermodynamic constraint imposed by Nernst's theorem, a more complete analysis of Luttinger's functional is required.

Consider the temperature derivative of $A_1(T)$ in Eq.~(\ref{eq:AKcorr}),
\begin{align} \label{eq:A1derivative}
\frac{dA_1(T)}{dT}
&=\frac1{ϖ_c}∫\liminff\f{dε}{2}\sin\f{ε}{ϖ_c}e^{-\f{\ImΣ^R_\dyn(ε,T)}{ϖ_c}}\notag\\
×&\brac{
-\frac{ε}{T}\!\para{\!\!\frac{∂\tanh\frac{ε}{2T}}{∂ε}\!\!}
-\frac1{ϖ_c}\tanh\frac{ε}{2T}\!\para{\!\!\frac{∂\ImΣ^R_\dyn(ε,T)}{∂T}\!\!}}\,.
\end{align}
The frequency integration in the first term in the square brackets vanishes in the limit of zero temperature because it is confined to an interval $∝k_BT$ about zero. This term is the only one present in the expression for the temperature derivative of $A_0(T)$. The second term is finite unless $∂\ImΣ^R_\dyn(ε,T→0)/∂{T}$ decays fast enough with $ε$. We emphasize that it is the behavior of $\ImΣ^R_\dyn(ε,T)$ at energies below as well as above $~k_BT$ that determines the slope of $A_1(T→0)$. We conclude that the violation of the Nernst's theorem constraint $dA_1(T→0)/d{T}=0$ in $Ω_{1}$ follows directly from the strong energy-dependence of $∂\ImΣ^R(ε,T→0)/∂{T}$ (see Appendix for specific example of such anomalous behavior). 

Analyticity of the Fermi liquid description of a normal metal ensures that oscillating components in the last two terms of Eq.~(\ref{eq:LF}) cancel out.\cite{Luttinger-QO} At the same time,  quasiparticle relaxation near the Fermi surface in Fermi liquids is weakly dependent on temperature and energy.\cite{Landau,Leggett} Therefore, analysis of $Ω_\osc(T,B)$ confined to $Ω_1$ alone is consistent in the Fermi liquid metal: $\ImΣ^R_\dyn(ε,T)$ has a weak temperature and frequency dependence. In contrast, both of these conditions---that of a weak temperature--and frequency--dependence of $\ImΣ^R_\dyn(ε,T)$ and that of analyticity which warrants cancellation of the last two terms in Eq.~(\ref{eq:LF})---break down  for correlated metals such as cuprates near the critical doping.\cite{Finkelstein-Luttinger, BychkovGorkov} To avoid violation of Nernst theorem, account of electronic correlations in magneto-scillations in these systems must include the effects captured by the other two terms in the Luttinger functional, Eq.~(\ref{eq:LF}).

\hypersetup{linkcolor=black,citecolor=lightblue,filecolor=black,urlcolor=black}
\begin{acknowledgments}
The work at the National High Magnetic Field Laboratory is supported by National Science Foundation Cooperative Agreement No. DMR-1157490 and the State of Florida. A.S. acknowledges the hospitality of the Aspen Center for Physics, where part of the work was done. The Aspen Center for Physics is supported by National Science Foundation Grant No. PHY-1066293. We thank A. Finkel'stein, N. Harrison, S. Kivelson, C.M. Varma for stimulating discussions.
\end{acknowledgments}

\appendix
\cleardoublepage

\section{Green's function representation of the thermodynamic potential in the Fermi gas.}

The first term in Eq.~(1) of the main text, $Ω_1=T∑_{iε_n,p}\ln{G}$ directly corresponds to the Green's function representation for the thermodynamic potential in the Fermi gas,\cite{Negele-Orland}
\begin{align}\label{eq:free1}
Ω_0 =& ∑_r T∑_{iε_n} e^{iε_n 0^+} \matzln\,G_r(iε_n) \notag\\
=& -∑_r ∫\f{dε}{4πi}  \tanh\f{ε}{2T}  \matzln\para{ε-E_r +μ}  e^{ε 0^+}  \,,
\end{align}
where $G_r(iε_n)=1/[iε_n-E_r+μ]$ is the Green's function and $E_r$ is a set of single-particle energies. The factor $e^{iε_n 0^+}$ ensures convergence of the integral on the second line at large negative values of $ε$. In the second line the sum over discreet frequencies has been transformed to a contour integral in the complex $ε$-plane. $\matzln(ε)=(1/2)[{\ln}_{→}(ε)+{\ln}_{←}(ε)]$, where $\ln_{→, ←}(ε)]$ is a logarithm function with brunch cut on the right and left side of the real axis respectively. Defined that way, it has a spectral weight $\Im\,\matzln(ε=x+i0)=-(π/2)\,\sign{ε}$.\cite{Negele-Orland} The integration contour in Eq.~(\ref{eq:free1}) consists of a line $ε=x+i0$ (immediately above the real axis) in the positive direction and the line $ε=x-i0$ in the negative $ε$-direction. Using the relation of $\Im\,\matzln(ε=x± i0)$ above and below the real axis we can leave only the part of the contour above the real axis,
\begin{align}\label{eq:free2}
Ω_0 =& -∑_r∫\liminff\f{dε}{2π}\tanh\f{ε}{2T}\Im\,\matzln\brac{ε-E_r +μ}e^{ε0^+}\,.
\end{align}
which evaluates to the well known expression for the thermodynamic potential of the non-interacting Fermi gas, $Ω_0=-T∑_r\ln\!\!\para{1+e^{\f{μ-E_r}{T}}}$. The contour of integration in Eq.~(\ref{eq:free2}) is the same as in Eq.~(2) of the main text. Eq.~(2) of the main text also omits the convergence factor $e^{ε 0^+}$.

\section{Summation over Landau level index.}

The sum over Landau level index in Eq.~(2) of the main text has the form
\begin{align}\label{eq:sum1}
∑_{λ_p} \Im\,\matzln(x - λ_p) \,,
\end{align}
where $λ_p=2πϖ_c(p+1/2)$,  $ϖ_c=ω_c/2π$, and $x = ϵ + μ - Σ^R(ϵ,T)$. We assume that $ϵ$ is slightly above the real axis, and therefore so is $x$. Because $\matzln(x)$ have the same algebraic properties as $\ln(x)$, we can rewrite this as a product
\begin{align}\label{eq:sum2}
\Im\,\matzln∏_{λ_p}(x-λ_p) \,.
\end{align}
The ultraviolet divergence in this expression can be regularized by modifying Eq.~(\ref{eq:sum2}) to account for the finite band width effects. However, as long as one is interested in the oscillating part of this expression (which originates from values of $λ_p$ close to $x$) the ultraviolet behavior is of no importance. Comparing with the the product representation of $\cos(x)$, Eq.~(\ref{eq:sum2}) can be written as
\begin{align}\label{eq:sum3}
\Im\,\matzln\cos\para{\frac{x}{2ϖ_c}} + ϖ_c ×\text{const}\,.
\end{align}
The first term in this expression leads to Eq.~(3) in the main text. Alternative way to arrive at the same answer is to  represent the sum $∑_p f(λ_p)$ via an integral in the complex $λ$-plane over a contour encircling all poles of $\tan\para{\sfrac{λ}{2ϖ_c}}$ on the real axis, \cite{Finkelstein-landau}
\begin{align}
-ϖ_c∑_{λ_p}f(λ_p)=∫\frac{dλ}{4πi}\,f(λ)\,\tan\f{λ}{2ϖ_c}\,,
\label{eq:sumlandau}
\end{align}
where
\begin{align} \label{eq:aaa}
f(λ) =&\Im\,\matzln[ϵ-λ+μ-Σ^R(ϵ,T)] \notag\\
=& -\f{π}2\,\sign[ϵ-λ+μ-Σ^R(ϵ,T)]
\end{align}

\section{Calculation of $A_1(T)$ in the Marginal Fermi liquid model.}

In this section we discuss  a specific example of an anomalous selfenergy that leads to a finite-temperature slope in $A_1(T→0)$.

Electronic transport measurements \cite{Fisher2014,Hussey2009,Mackenzie2013} suggest that the quasiparticle self-energy in high temperature superconductors is linear in temperature in the zero-energy limit. ARPES measurements suggest that $\ImΣ^R(ε,T)$ is linear in energy in the opposite limit, $ε≫T$.\cite{ARPES-Valla,ARPES-ZX} The marginal Fermi liquid model captures both of these observations with the hypothesis that electrons near the Fermi surface interact with a local (momentum--independent) bosonic mode which has constant (as a function of frequency) spectral weight at zero temperature.\cite{Varma1989,AjiVarma,ShekhterVarma,Senthill2005} At finite temperature the spectral weight is suppressed at low frequencies, $ℏ ω≲ k_BT$. Specifically, the spectral weight of the fluctuation mode has a form
\begin{align}
\Im\Q^R(ω,T) = -\tanh{\!\frac{ζω}{2T}},\quad  |ω|≲{Λ}\,.
\label{eq:Q}
\end{align}
where $Λ$ is the ultraviolet cutoff (about $0.4eV$ in cuprates\cite{ZhuVarma}) and $ζ$ is a numeric factor of order unity\cite{AjiVarma,zeta-comment}.

\begin{figure}[ht!]
\centerline{\includegraphics[width=\columnwidth]{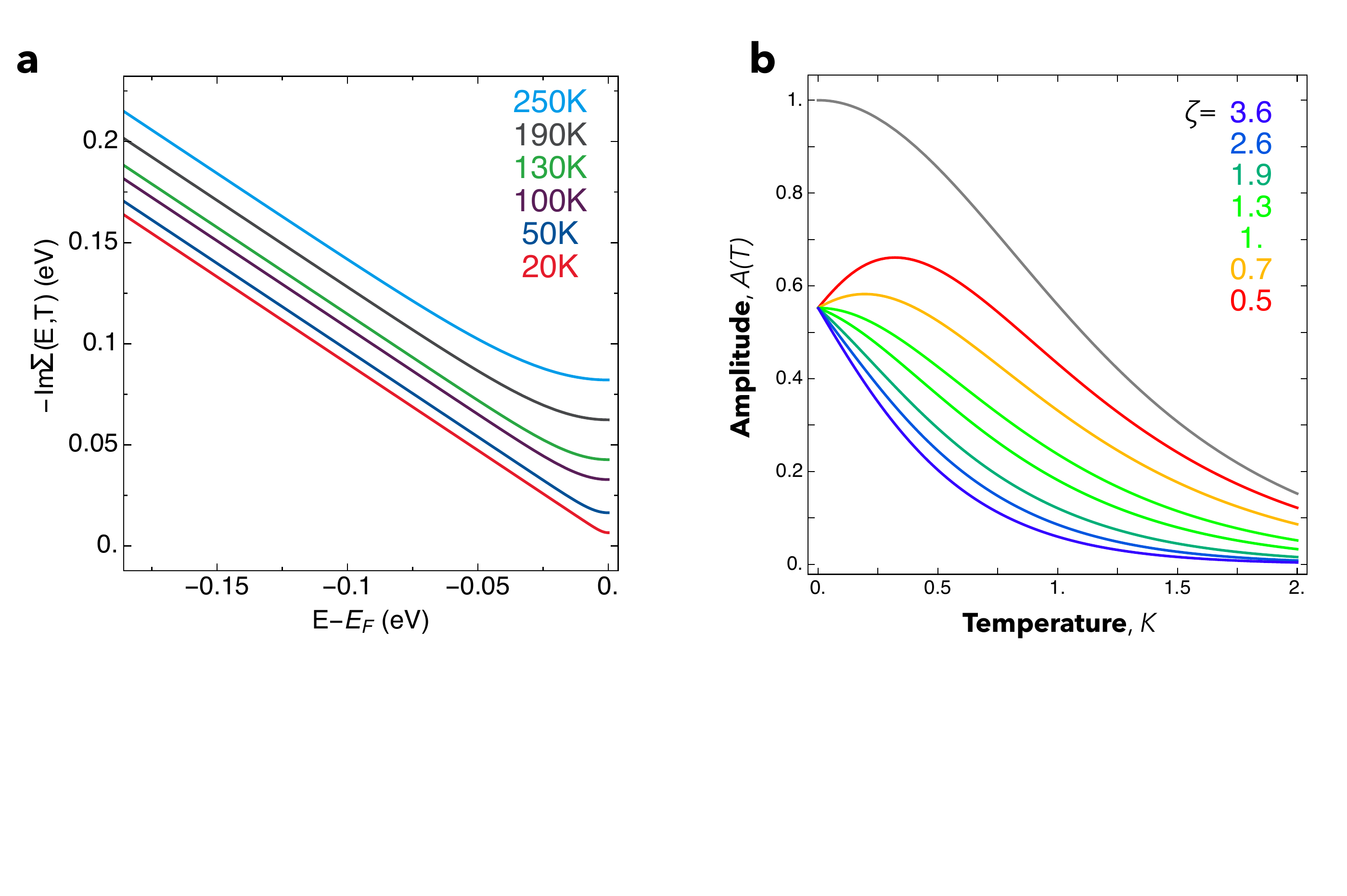}}
\caption{Anomalous self-energy and its effect on the amplitude of quantum oscillations. \textbf{a} Energy and temperature dependence of the imaginary part of the anomalous self-energy in the marginal Fermi liquid, Eq.~(\ref{eq:ImSig}). \textbf{b} Amplitude of quantum oscillations vs temperature for several values of the numeric parameter $ζ$ (for $g_s=1$). Gray line indicates the $A_0(T)$ form of temperature dependence (i.e., $g_s=0$ ).
 }
\label{fig:new}
\end{figure}

The selfenergy resulting from interaction with such a mode has a form
\begin{align}
\ImΣ^R(ε,T)
 =& g_s\!\!∫\liminff\f{dω}2\!\!\brac{\coth\!\f{ω}{2T} - \tanh\f{ω-ε}{2T}}\!\!\Im\Q^R(ω,T)\,,
\label{eq:ImSig}
\end{align}
where $g_s$ is the coupling constant  (see Section \ref{sec:selfenergy} for details). The energy and temperature dependence of $\ImΣ^R(ε,T)$ is shown in Fig.~(\ref{fig:new}a) for $ζ=5$. $\ImΣ^R(ε,T)$ is linear-in-temperature both at low energy, $\ImΣ^R(ε≪T,T)=-g_s α^0(ζ)T$ and, importantly, at high energy as well, $\ImΣ^R(ε≫T,T)=-g_s[ε+α^1(ζ)T]$, where  $α^{0,1}(ζ)$ are $ζ$-dependent numeric coefficients. The finite value of $α^1(ζ)=∂\ImΣ^R(ε≫T,T→0)/∂{T}$  implies a finite value of the second term in Eq.~(\ref{eq:A1derivative}) and therefore a finite value of $dA_1(T→0)/dT$, Fig.~(\ref{fig:new}b). This illustrates breakdown of the Nernst theorem in $Ω_{1,\osc}$ for a general marginal Fermi liquid model ($ζ ≠ 1$).

\subsection{Calculation of electron selfenergy in marginal Femi liquid}
\label{sec:selfenergy}

Inelastic quasiparticle relaxation rate in the marginal Fermi liquid, $1/τ_\dyn= -2\ImΣ^R_\dyn(ϵ,T)$, is determined by the self-energy diagram,
\begin{align}\label{eq:first}
Σ(iϵ_n) = & \tadpole{iϵ;p}{iω;p-p'}{iϵ-iω;p'}{iϵ;p}\notag\\
=& -g_0T∑_{iω_n}∫\frac{d^2p'}{ν_0(2π)^2} G(iϵ_n-iω_n)\Q(iω_n)
\end{align}
Writing frequency sum as a contour integral in the complex-$ϵ$ plane and continuing analytically to the real-$ϵ$ axis,\cite{Negele-Orland}
\begin{align}
&\tadpole{iϵ}{iω}{iϵ-iω}{iϵ}
\xrightarrow{iϵ_n>0}
-g_0∫\frac{dω}{4πi} ∫\frac{d^2p'}{ν_0(2π)^2}×\notag\\
&×\Big[
\coth\frac{ω}{2T} \;\tadpole{R}{R-A}{R}{R}
-\tanh\frac{ω-ϵ}{2T} \;\tadpole{R}{R}{R-A}{R} \Big]\,,
\end{align}
(here $R,A$ represent retarded and advanced Green's functions respectively) we obtain~:
\begin{align}\label{eq:second}
\ImΣ^R(ϵ,p) =&
-g_0 ∫\liminff\frac{dω}{2π}∫\frac{d^2p'}{ν_0(2π)^2}
[\coth\frac{ω}{2T}-\tanh\frac{ω-ϵ}{2T}] \notag\\
× & \Im{G}^R(ϵ-ω,p')\Im\Q(ω)\,.
\end{align}
Locality ($q$-independence) of $\Q(ω)$ allows us to perform momentum integration using an identity
\begin{align}\label{eq:aaa}
∫\frac{d^2p'}{(2π)^2} \Im G^R(ϵ, p)
= ν_0∫_{-∞}^{∞} dξ\Im{G}^R(ϵ,ξ)
=-ν_0π \,,
\end{align}
where $ν_0$ is the density of states of the two-dimensional electron gas near the Fermi surface. Eq.~(\ref{eq:second}) takes the form
\begin{align}\label{eq:Im-Sigma-general}
\ImΣ^R(ϵ) =& g_0 \Im\SE(ϵ)  \notag \\
\Im\SE(ϵ) =&∫\liminff{dω}\frac12 [\coth\frac{ω}{2T}-\tanh\frac{ω-ϵ}{2T}]
\;\Im\Q^R(ω) \,.
\end{align}
Specifically, for the marginal Fermi liquid form of the fluctuation spectrum, Eq.~(\ref{eq:Q}), we obtain (for $ϵ≪Λ$)
\begin{align}\label{eq:Im-Sigma-1}
\Im\SE^R(ϵ) =& -∫\liminff{dω}\frac12[\coth\frac{ω}{2T}-\tanh\frac{ω-ϵ}{2T}]
\;\tanh\frac{ζω}{2T}\,.
\end{align}
The right-hand side is linear in temperature at small energies and is linear in energy at small temperature. Importantly, it also has a finite linear-in-temperature component, even at high energy $|ϵ|≫ T$. Specifically,
\begin{align}
-\Im\SE^R(ϵ)_T
=&  α^0(ζ)T  \,,  & |ϵ|≪T    \notag\\
=&  |ϵ| + α^1(ζ)T \,,  & |ϵ|≫T
\end{align}
where $α^{0,1}(ζ)$ are numerical factors of order unity defined as
\begin{align}\label{eq:alphas}
α^{0}(ζ) = -\frac{d\Im\SE(ϵ)}{dT}\Big|_{ϵ→0}
=& ∫\liminff{dω}\frac{\tanh\mfrac{ζω}{2}}{\sinhω}    \notag\\
α^{1}(ζ) = -\frac{d\Im\SE(ϵ)}{dT}\Big|_{ϵ≫ T}
=& ∫\liminff{dω}\frac{e^{ω/2}\sinh\mfrac{[ζ-1]ω}{2}}{\cosh\mfrac{ζω}{2}\sinhω}\,.
\end{align}
The self-energy in Eq.~(\ref{eq:Im-Sigma-1}) exhibits competition of the dominant energy scales,  $\Im\SE^R(ϵ) ∼ \max[ϵ, α^0(ζ)T]$ as is expected on general grounds near critical point. It also shows that in this model the dependence on a subdominant variable, temperature in this case, is also strong, linear-in-T with a different slope.

The real part can be obtained from Kramers-Kronig relations applied to  $\SE^R(ϵ)$
\begin{align}\label{eq:KramersKronig}
	\Re f(ω) = \HT{\Im f(ω)}
	≡ \frac1{π}\;P\!\!\!\!\!\!∫ dω'\frac{\Im f(ω')}{ω'-ω}\,.
\end{align}
This integral is formally divergent for $f(ϵ)$ from Eq.~(\ref{eq:Im-Sigma-1}). The divergence can be traced to the zero-temperature expression. To isolate this divergence we note that the finite-temperature expression for $\Im\SE^R(ϵ)$ can be approximately represented as a convolution of $C_{ζ}(s)$ with the zero-temperature expression for $\Im\SE^R(ϵ,T=0)$,
\begin{align}\label{eq:smearing}
\Im\SE^R(ϵ)_T =&  -α^1(ζ)T + ∫\liminff{ds}\; C_ζ(s)
\Im\SE^R(ϵ-s)_{T=0}  \,,
\end{align}
where
\begin{align}
C_ζ(s)=\frac{d\tanh\para{\frac{p_{ζ}s}{2T}} }{2ds}\,
\end{align}
where $p_{ζ}= 2\ln2/[α^0(ζ)-α^{1}(ζ)]$ and $α^{0,1}(ζ)$ are defined in Eq.~(\ref{eq:alphas}).\cite{SF-LinearT-prb}. The same convolution representation will apply for the real part at finite temperatures,
\begin{align}
\Re\SE^R(ϵ)_T =&∫\liminff{ds}C_ζ(s)\Re\SE^R(ϵ-s)_{T=0}\,.
\end{align}
This allows to isolate the divergence associated with the Kramers-Kronig integrals to the zero-temperature expressions in which it can be traced to the behavior near ultra-violet cutoff $Λ$:
\begin{align}\label{eq:ImSigmaZeroT}
&\Im \SE^R(ϵ)_{T=0} = ∫_0^{ϵ} dω\Im\Q^R(ω)_{T=0}
= -\min(|ϵ|, Λ) \notag\\
&\Re\SE^R(ϵ)_{T=0}=\HT{\Im \SE(ϵ)_{T=0}}  \notag\\
=&  -\frac2{π}
\Big[\!ϵ\ln\!\frac{Λ}{ϵ}\!+\!
\frac12(ϵ\!+\!Λ)\ln\!\frac{ϵ\!+\!Λ}{Λ}\!+\!\frac12(ϵ\!-\!Λ)\ln\!\frac{ϵ\!-\!Λ}{Λ}\!\Big]\notag\\
&\xrightarrow{ϵ≪Λ} -\frac2{π}\brac{ϵ\ln\frac{Λ}{|ϵ|}+ϵ}\,.
\end{align}
We finally obtain~:
\begin{align}
\Re\SE^R(ϵ≪Λ_0)_T
≈& -\!\!\!∫\liminff{ds}\, C_ζ(s)\; \frac2{π}\brac{(ϵ-s)\ln\frac{eΛ_0}{|ϵ-s|}}\,.
\end{align}

\end{document}